\documentclass[twocolumn,aps,showpacs,superscriptaddress]{revtex4}
\usepackage{amssymb}
\usepackage{amsfonts}
\usepackage{amsmath}
\usepackage{graphicx}
\usepackage{bm}
\usepackage{multirow}

\newcommand{\vect}{\boldsymbol}
\begin{document}

\title{Generic Miniband Structure of Graphene on a Hexagonal Substrate}

\author{J.~R.~Wallbank}
\affiliation{Department of Physics, Lancaster University, Lancaster, LA1 4YB, UK}
\author{A.~A.~Patel}
\affiliation{Department of Physics, Lancaster University, Lancaster, LA1 4YB, UK}
\affiliation{Department of Physics, Indian Institute of Technology Kanpur, Kanpur 208016, India}

\author{M.~Mucha-Kruczy\'{n}ski}
\affiliation{Department of Physics, Lancaster University, Lancaster, LA1 4YB, UK}

\author{A.~K.~Geim}
\affiliation{Centre for Mesoscience and Nanotechnology, University of Manchester, Manchester M13 9PL, UK}

\author{V.~I.~Fal'ko}
\affiliation{Department of Physics, Lancaster University, Lancaster, LA1 4YB, UK}
\affiliation{DPMC, University of Geneva, 24 Quai Ernest-Ansermet, CH1211 Gen\'eve 4, Switzerland}

\date{\today}

\begin{abstract}
Using a general symmetry-based approach, we provide a classification of generic miniband structures for electrons in graphene placed on substrates with the hexagonal Bravais symmetry. In particular, we identify conditions at which the first moir\'e miniband is separated from the rest of the spectrum by either one or a group of three isolated mini Dirac points and is not obscured by dispersion surfaces coming from other minibands. In such cases the Hall coefficient exhibits two distinct alternations of its sign as a function of charge carrier density.
\end{abstract}


\maketitle

Recently, it has been demonstrated that the electronic quality of graphene-based devices can be dramatically improved by placing graphene on an atomically flat crystal surface, such as hexagonal boron nitride (hBN) \cite{dean_natnano_2010,xue_natmat_2011,bresnehan_acsnano_2012,decker_nanolett_2011,mayorov_nanolett_2011,kim_apl_2011,wang_ieee_2011}. At the same time, graphene's electronic spectrum also becomes modified, acquiring a complex, energy-dependent form caused by incommensurability between the graphene and substrate crystal lattices \cite{yankowitz_natphys_2012,ortix_prb_2012,kindermann_prb_2012}. In particular, for graphene placed on hBN, the difference between their lattice constants and crystallographic misalignment generate a hexagonal periodic structure known as a moir\'e pattern \cite{xue_natmat_2011,decker_nanolett_2011,yankowitz_natphys_2012,ortix_prb_2012,kindermann_prb_2012}. 
The resulting periodic perturbation, usually referred to as a superlattice, acts on graphene's charge carriers and leads to multiple minibands and the generation of secondary Dirac-like spectra. The resulting new Dirac fermions present yet another case where graphene allows mimicking of QED phenomena under conditions that cannot be achieved in particle physics experiments. In contrast to relativistic particles in free space, the properties of secondary Dirac fermions in graphene can be affected by a periodic sublattice symmetry breaking and modulation of carbon-carbon hopping amplitudes, in addition to a simple potential modulation. The combination of different features in the modulation results in a multiplicity of possible outcomes for the moir\'e miniband spectrum in graphene which we systematically investigate in this article.

To describe the effect of a substrate on electrons in graphene at a distance, $d$, much larger than the spacing, $a$, between carbon atoms in graphene's honeycomb lattice, we use the earlier observation  \cite{santos_prl_2007,bistritzer_prb_2010,bistritzer_prb_2011,santos_prb_2012,yankowitz_natphys_2012,ortix_prb_2012,kindermann_prb_2012} that, at $d\gg a$ the lateral variation of the wavefunctions of the p$^\text{z}$ carbon orbitals is smooth on the scale of $a$. This is manifested in the comparable sizes of the skew and vertical hopping in graphite and permits an elegant continuum-model description \cite{santos_prl_2007,bistritzer_prb_2010,bistritzer_prb_2011,santos_prb_2012} of the interlayer coupling in twisted bilayers and the resulting band structure. A similar idea applied to graphene on a hBN substrate \cite{yankowitz_natphys_2012,ortix_prb_2012,kindermann_prb_2012} suggests that a substrate perturbation for Dirac electrons in graphene can be described in terms of simple harmonic functions 
corresponding to the six smallest reciprocal lattice vectors of the moir\'e superlattice.

%%%%%%%% Figure: fig1  %%%%%%%%%%%%%%%%%%%%
	\begin{figure*}[htbp]
	\centering
	\includegraphics[width=.92\textwidth]{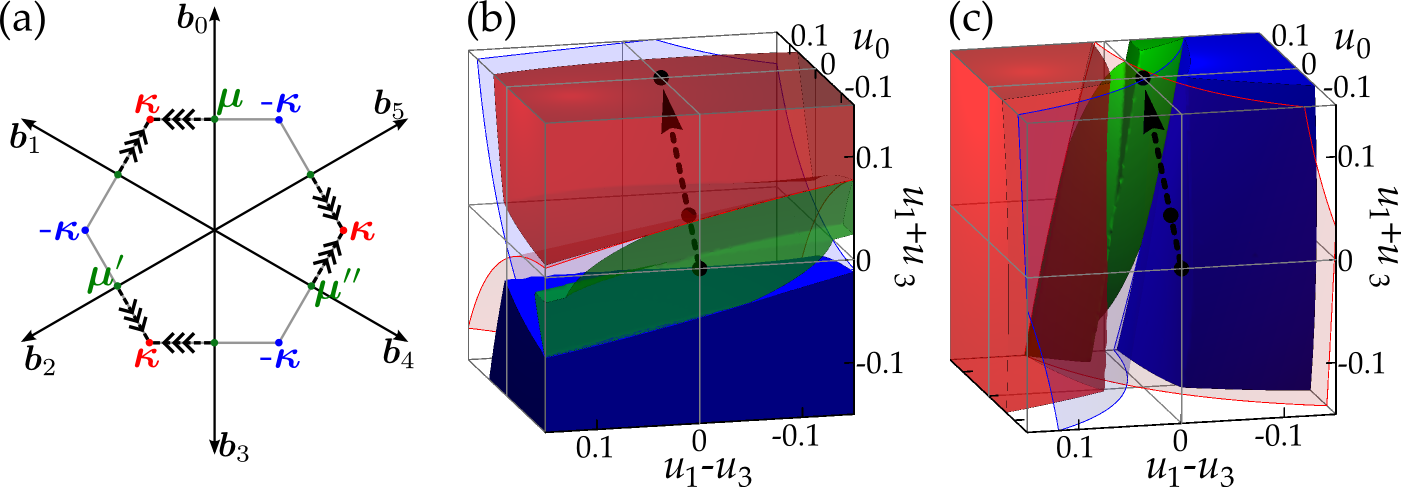}
	\caption{
	(a) The hexagonal Brillouin zone for the moir\'e superlattice. 
	(b) Three volumes in the space of the moir\'e superlattice parameters where the edge of the first miniband, in graphene's valence band, contains an isolated mDP at the $\vect\kappa$-point (red) or the $-\vect\kappa$-point (blue) or three isolated mDPs at the sBZ edge (green). Parameters for which the $\pm\vect\kappa$-point is triple degenerate are shown by the red and blue surfaces. The black dots represent sets of perturbation parameters for which miniband spectra are shown in Fig.~\ref{fig:fig2}.
	(c)The same for the conduction band in graphene.
	}
	\label{fig:fig1}
	\end {figure*}

Below, we shall use a similar approach to analyse the generic properties of moir\'e minibands for electrons in graphene subjected to a substrate with a hexagonal Bravais lattice with a slightly different lattice constant of $(1\!+\!\delta)\sqrt{ 3} a$, $|\delta|\!\ll\!1$, compared to that of $\sqrt 3a$ for graphene, and a small misalignment angle, $\theta\!\ll\!1$.
The moir\'e pattern harmonics are described by vectors 
\begin{align}
 &\vect b_{m=0,\cdots5}\!=\!\hat R_{\frac{2\pi m}{6}} \vect b_0,\;\;\;\vect b_0\!=\!\left[1\!-\!(1\!+\!\delta)^{-1}\hat R_\theta\right]\! \left(\!0,\!\frac{4 \pi}{3 a}\!\right)\!,
\end{align}
with length $|\vect b_0|\!\equiv \!b\!\approx\!\frac{4\pi}{3a}\sqrt{\delta^2\!+\!\theta^2}$, which can be obtained from each other by the anticlockwise rotation, $\hat R_{2\pi m/6}$.  
For a substrate with a simple hexagonal lattice or a honeycomb lattice with two identical atoms, the perturbation created for graphene electrons is inversion-symmetric. For a honeycomb substrate where one of the atoms would affect graphene electrons stronger than the other (e.g.~such as hBN, for which the occupancy and size of the p$^\text{z}$ orbitals are different) the moir\'e potential can be modelled as a combination of a dominant inversion-symmetric part with the addition of a small inversion-asymmetric perturbation,
\begin{widetext}
  \begin{align}
  \hat H=v\vect p \cdot \vect\sigma&+u_0v bf_1(\vect r)+u_3v b f_2(\vect r)\sigma_3\tau_3+u_1v \left[\vect l_z\times\nabla f_2(\vect r)\right]\cdot\vect \sigma\tau_3+  u_2 v \nabla f_2(\vect r)\cdot\vect \sigma\tau_3 \label{eq:H} \\
 &+\tilde u_0 v b f_2(\vect r)+\tilde u_3 v b f_1(\vect r)\sigma_3\tau_3+   \tilde u_1 v  \left[\vect l_z\times\nabla f_1(\vect r)\right]\cdot\vect \sigma\tau_3+\tilde u_2 v  \nabla f_1(\vect r)\cdot\vect \sigma\tau_3. \nonumber
 \end{align}
\end{widetext}

%%%%%%%%%%%%%%%% Hamiltonian description %%%%%%%%%%%%%%%%%%%%%%%%%%%
The Hamiltonian, $\hat H$, acts on four-component wavefunctions, $(\Psi_{AK},\Psi_{BK},\Psi_{BK'},-\Psi_{AK'})^T$, describing the electron amplitudes on graphene sublattices $A$ and $B$ and in two principal valleys, $K$ and $K'$. It is written in terms of direct products $\sigma_i\tau_j$, of Pauli matrices $\sigma_i$ and $\tau_j$ separately acting on sublattice and valley indices.  
The first term in $\hat H$ is the Dirac part, with $\vect p\!=\!-i\nabla\!-\!e\vect A$ describing the momentum relative to the centre of the corresponding valley, with $\nabla \!\times \!\vect A\!=\!B$. The rest of the first line in Eq.~\eqref{eq:H} describes the inversion-symmetric part of the moir\'e perturbation, whereas the second line takes into account its inversion-asymmetric part.
In the first line, the first term, with $f_1(\vect r)=\sum_{m=0...5}e^{i\vect b_m\cdot\vect r}$, describes a simple potential modulation. 
The second term, with $f_2(\vect r)=i\sum_{m=0...5}(-1)^me^{i\vect b_m\cdot\vect r}$, accounts for the $A$-$B$ sublattice asymmetry, locally imposed by the substrate. 
The third term, with unit vector $\vect l_z$, describes the influence of the substrate on the $A$-$B$ hopping: consequently \cite{Iordanskii85,Foster2006,Morpurgo2006}, this term can be associated with a pseudo-magnetic field, $e\beta=\pm u_1 b^2f_2(\vect r)$, which has opposite signs in valleys $K$ and $K'$. 
Each of the coefficients $|u_i|\ll1$ in Eq.~\eqref{eq:H} is a dimensionless phenomenological parameter with the energy scale set by $v b\approx2\pi\sqrt{\delta^2+\theta^2}\gamma_0$, where $\gamma_0\approx3\,$eV is the nearest neighbour hopping integral in the Slonczewski-Weiss tight binding model \cite{Slonczewski_APS_1958}. 
Concerning the inversion-asymmetric part, the second line in Eq.~\eqref{eq:H}, we assume that $|\tilde u_i| \ll |u_i|$.
Note that the last term in each line can be gauged away using $\vect\psi\rightarrow e^{-i \tau_3(u_2   f_2+\tilde u_2  f_1) } \vect\psi$. 

%%%%%%%%%%%%%%%%%%%%%%%%%% Microscopic models %%%%%%%%%%%%%%%%%%%%%%%%%
Hamiltonian, $\hat H$, may be used to parametrise any microscopic model compatible with the symmetries of the system (see Appendix 1) and the  dominance of the simplest moir\'e harmonics, $e^{i\vect b_m\cdot\vect r}$, in the superlattice perturbation. The values that parameters $u_i$ take are listed in Table \ref{table_1} for several models of graphene on an hBN substrate, both taken from the recent literature \cite{yankowitz_natphys_2012,ortix_prb_2012,kindermann_prb_2012} and analysed in Appendix 2, including a simple model in which the hBN substrate is treated as a lattice of positively charged nitrogen nuclei with a compensating homogeneous background of electron P$^{\text{z}}$ orbitals. The examples of model-dependent values of parameters $u_i$, listed in Table  \ref{table_1}, indicate that the combination of several factors can strongly shift the resulting moir\'e perturbation across the parameter space in Fig. \ref{fig:fig1}.
That is why, in this work, we analyse the generic features of the miniband spectra generated by the moir\'e superlattice, rather than attempt to make a brave prediction about its exact form for a particular substrate.
\begin{center}
\begin{table}[tbhp]
\begin{tabular}{|l|c|c|c|c|}\hline
  \textbf{Model}  & $vbu_0$  & $vbu_1$ & $vbu_2$ & $vbu_3$  \\
 &[meV]&[meV]&[meV]&[meV] \\ \hline
Potential modulation \cite{yankowitz_natphys_2012}   & $60$ & $0$ & $0$ & $0$ \\\hline
2D charge modulation \cite{ortix_prb_2012} &  $-\frac{V_0}{2}$ &  $0$ & $0$ & $\frac{\sqrt3 V_0}{2}$\\\hline
One-site version of G-hBN& \multirow{2}{*}{$1.6$} & \multirow{2}{*}{$\frac{-3.2\delta}{\sqrt{\delta^2+\theta^2}}$} &\multirow{2}{*}{ $\frac{3.2\theta}{\sqrt{\delta^2+\theta^2}}$} &\multirow{2}{*}{$-2.8$} \\
  hopping \cite{kindermann_prb_2012}  (Appendix 2b)& & & & \\\hline
Point charge lattice  & \multirow{2}{*}{$\frac{\tilde v}{2}$} & \multirow{2}{*}{$\frac{-\tilde v\delta}{\sqrt{\delta^2+\theta^2}}$} &  \multirow{2}{*}{$\frac{\tilde v\theta}{\sqrt{\delta^2+\theta^2}}$} &\multirow{2}{*}{$-\frac{\sqrt 3 \tilde v}{2}$}\\  
(Appendix 2a), $0.6\!\leq\!\tilde v\!\leq\!3.4$ & & & & \\\hline
\end{tabular}
\caption{The inversion-symmetric parameters, $vbu_i$, for various  models of the moir\'e superlattice.  
In the 2D charge modulation model \cite{ortix_prb_2012}, $V_0$ is a phenomenological parameter. The G-hBN hopping model in Ref. \cite{kindermann_prb_2012} used the hopping parameter from twisted bilayer graphene.
Estimates in Appendix 2 show that the sets of parameters using a model of point charges attributed to nitrogen sites and for the G-hBN hopping model are very similar.
}
\label{table_1}
\end{table}
\end{center}

%%%%%%%%%%%%%%%%%%%%%%%%%%%%%% Symmetries %%%%%%%%%%%%%%%%%%%%%%%%%%%%%%%%%
In the absence of a magnetic field, the Hamiltonian Eq.~\eqref{eq:H} obeys time-reversal symmetry, which follows from both $\sigma_i$ and $\tau_i$ changing sign upon the transformation $t\rightarrow-t$ \cite{Aleiner_prl_2006}. As a result, $\epsilon_{\vect K+\vect p}=\epsilon_{\vect K'-\vect p}$  and we limit the discussion of minibands to the $K$ valley. Subject to this limitation the bandstructure for the inversion-symmetric superlattice perturbation obeys the $c_{3v}$ symmetry. Moreover, using the commutation properties of $\sigma_i$ one can establish that
  \begin{align}
\!\epsilon^{u_0,u_1,u_3}_{\vect K+\vect p}\!&=\!-\epsilon^{-u_0,-\!u_1,u_3}_{\vect K-\vect p}\!=\!-\epsilon^{-u_0,u_1,-\!u_3}_{\vect K+\vect p}\!=\!\epsilon^{u_0,-\!u_1,-\!u_3}_{\vect K-\vect p}\!.\label{eq:sym}
\end{align}

%%%%%%% Figure: fig2  %%%%%%%%%%%%%%%%%%%%
	\begin{figure*}[htbp]
	\centering
	\includegraphics[width=.92\textwidth]{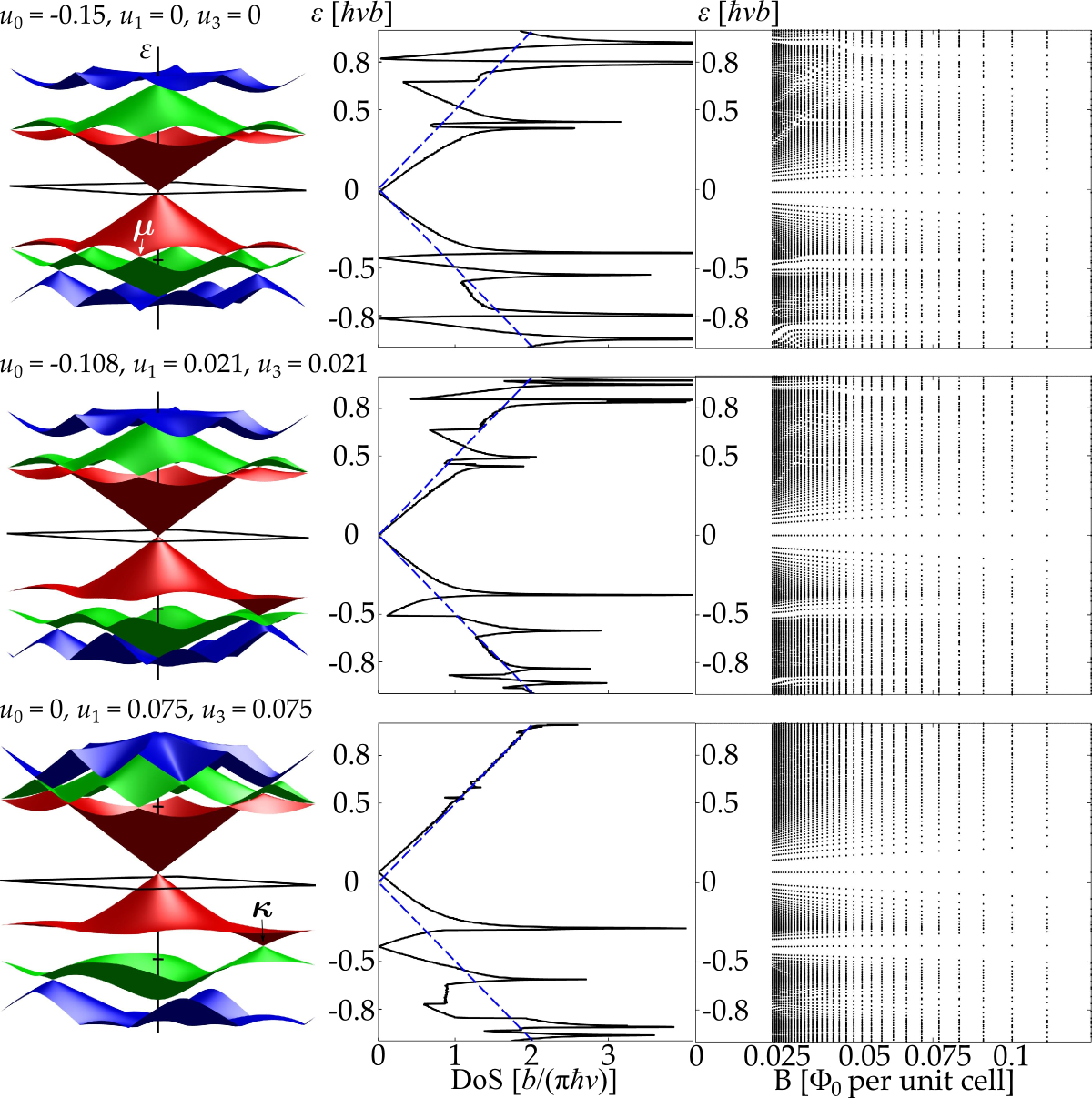}
	\caption{
Numerically calculated moir\'e miniband (left), the corresponding density of states (centre), and Landau level spectrum (right) for electrons in the vicinity of graphene's $K$ point. Here we use the rhombic sBZ, so that the $c_{3v}$ symmetry of the moir\'e superlattice spectrum is not obviously seen in the images.}
	\label{fig:fig2}
	\end {figure*}

%%%%%%%%%% calculation method %%%%%%%%%%%%%%%%%%%%% 
To calculate the miniband spectrum for $\hat H$ in Eq.~\eqref{eq:H} we perform zone folding (in the graphene $K$ valley) bringing states with momenta related by the reciprocal lattice vectors $n_1\vect b_1+n_2\vect b_2$ of the moir\'e pattern to the same point of the superlattice Brillouin zone (sBZ) in Fig.~\ref{fig:fig1}(a). Then, we calculate the matrix elements of $\hat H$ between those states and diagonalise the corresponding Heisenberg matrix numerically exploring the parametric space $(u_0,u_1,u_3)$ of the dominant inversion-symmetric part of the moir\'{e} perturbation shown in Fig \ref{fig:fig1} (b,c). The size of the matrix is chosen to guarantee the convergence of the calculated energies for the three lowest minibands in both the conduction band ($s=+1$) and the valence band ($s=-1$). 
Below, we discuss the generic features of the moir\'e miniband spectra for the characteristic points in the parametric space $(u_0,u_1,u_3)$, marked using black dots in Fig.~\ref{fig:fig1}(b,c), using both the numerically calculated dispersion surfaces in Fig. \ref{fig:fig2} and analytical perturbation theory analysis.

%%%%%%%%%%%%%%%%%% zero energy DP %%%%%%%%%%%%%%%%%%%%%%%
For the  zero-energy Dirac point in graphene, there are only the original $\vect p=0$ states in each valley that appear at $\epsilon=0$ upon zone folding.
For all three characteristic spectra shown in Fig.~\ref{fig:fig2}, for the inversion-symmetric moir\'e perturbation, the gapless Dirac spectrum persists at low energies near the conduction-valence band edge with almost unchanged Dirac velocity, $\left[1+O(u^2)\right]v$. The inversion-asymmetric terms $\tilde{u}_{i}$ are able \cite{kindermann_prb_2012} to open a minigap at the Dirac point.

%%%%%%%%%%%%%% mu point %%%%%%%%%%%%%%%%%%
 For the point $\vect \mu=\vect b_0/2$ on the edge of the first sBZ, zone folding brings together two degenerate plane wave states, $|\vect \mu+\vect q\rangle$ and $|\vect \mu+\vect b_3+\vect q\rangle$. The splitting of these degenerate states by the moir\'e potential in Eq.~\eqref{eq:H} can be studied using degenerate perturbation theory. The corresponding $2\times 2$ matrix, expanded in small deviation $\vect q$ of the electron momentum from each of the three sBZ $\vect \mu$-points \cite{footnote_mu_points} has the form
     \begin{align}
  &\hat H_{\vect\mu+\vect q}= vb\begin{pmatrix}
           E_{\vect \mu} +s \frac{q_y}{b} & H_{12}\\
           H_{12}^* & E_{\vect \mu}-s \frac{q_y}{b}\\           
           \end{pmatrix}, \label{eq:mu}\\
 &E_{\vect \mu}\approx  \frac{s}{2}+\frac{sq_x^2}{b^2}    ,\nonumber \\
&H_{12}\approx   (su_1-u_3)-i(s\tilde u_1- \tilde u_3)+2 \frac{q_x}{b} (u_0+i \tilde u_0). \nonumber  
\end{align}
For the inversion-symmetric perturbation, the dispersion relation resulting from Eq.~\eqref{eq:mu} contains an anisotropic mini Dirac point (mDP) \cite{yankowitz_natphys_2012,park_prl_2008,guinea_ptrsa_2010} with Dirac velocity component $\approx 2u_0 v$ in the direction of the sBZ edge and $\approx v$ in the perpendicular direction.
This feature is clearly seen at the $\vect \mu$-point of the first moir\'e miniband in the valence band, in the top row of Fig.~\ref{fig:fig2}. Note that the electron spectrum is not symmetric between the valence and conduction bands and that the mDPs at the $\vect \mu$-point in the conduction band are obscured by an overlapping spectral branch.

Moving in parameter space, e.g., along the line shown in Fig.~\ref{fig:fig1}(b), the positions of the three anisotropic mDPs shift along the sBZ edge towards the sBZ corners: either $\vect \kappa=(\vect b_4+\vect b_5)/3$, or $-\vect \kappa$, as shown by arrowed lines in Fig.~\ref{fig:fig1}(a). In general, a spectrum with three isolated mDPs at the sBZ edge is typical for the green volume in the parameter space in Fig.~\ref{fig:fig1}(b) for the valence band, or Fig.~\ref{fig:fig1}(c) for the conduction band.
In contrast, for $(u_0,u_1,u_3)$ in the clear part of the parameter space, mDPs on the edge of the first sBZ are overshadowed by an overlapping spectral branch, as is the case on the conduction band side for all three cases shown in Fig.~\ref{fig:fig2}. 

For the points in Fig.~\ref{fig:fig1}(b,c) on the red and blue surfaces, the three mDPs reach the $\vect \kappa$-point, forming a triple degenerate band crossing, as in the valence band spectrum shown in the middle row of Fig.~\ref{fig:fig2}, which can be traced using the perturbation theory analysis of the band crossing at $\vect \kappa$ discussed below.

The third line in Fig.~\ref{fig:fig2} shows the third type of spectrum of moir\'e minibands, characteristic for the red and blue volumes of the parameter space in Fig.~\ref{fig:fig1}. The characteristic feature of such spectra consists in a single isolated mDP, at the $\pm\vect\kappa$-point, in the valence band (Fig.~\ref{fig:fig1}(b)) or the conduction band (Fig.~\ref{fig:fig1}(c)).

 For the $\vect \kappa$ and $-\vect \kappa$-points, zone folding brings together three degenerate plane wave states,  $|\zeta( \vect\kappa +\vect q)\rangle$, $|\zeta( \vect\kappa +\vect b_1+\vect q)\rangle$, and $|\zeta( \vect\kappa+\vect b_2 +\vect q)\rangle$ (where $\zeta=\pm$), whose splitting is determined by 
%  
%%%%%%%%%%%%%%%%%%%%%% kappa point %%%%%%%%%%%%%%%%%%%%%%
\begin{align}
 &\hat H_{\zeta(\vect\kappa+\vect q)}\!=\! vb\!\begin{pmatrix} 
  \frac{s}{\sqrt{3} }\!+\!\frac{sq_x}{b}      &w_{\zeta}	 & w_{\zeta}^*\\
w_{\zeta}^*	& \frac{s}{\sqrt{3} }\!-\!s\frac{q_x-\sqrt 3 q_y}{2b}    	&-w_{\zeta}     \\
w_{\zeta}   				& -  w_{\zeta}^*         	 &\frac{s}{\sqrt{3} }\!-\!s\frac{q_x+\sqrt 3 q_y}{2b}\\
 \end{pmatrix}\!\!,\nonumber\\
 &w_{\!\zeta}\!\approx\!\frac{1}{2}\!\!\left[\! \left(\!u_0  \!\!-\!\!2s\zeta u_1\!+\!\sqrt 3 \zeta u_3\!\right)\!\!+\!i\zeta\!\left(\! \tilde u_0 \!+\!2 s\zeta\tilde u_1\!\!-\!\!\sqrt3\zeta\tilde u_3 \!    \right)  \! \right]\!.\!\label{eq:kappa}
 \end{align}
For $w_\zeta\neq0$, the inversion-symmetric terms in $\hat H_{\zeta(\vect\kappa+\vect q)}$ partially lift the $\zeta\vect \kappa$-point degeneracy into a singlet with energy $(\frac{s}{\sqrt 3}-2w_{\zeta} )vb$ and a doublet with energies $(\frac{s}{\sqrt 3}+w_{\zeta} )vb$, so that a distinctive mDP \cite{guinea_ptrsa_2010} characterised by Dirac velocity $v_{\vect\kappa}=\left[1+O(u)\right]\frac{v}{2}$ \cite{ortix_prb_2012} is always present at  $\pm\vect \kappa$ somewhere in the spectrum \cite{footnote_kappa}. 
This behaviour reflects the generic properties of the symmetry group of wave vector $\vect \kappa$ which has the two-dimensional irreducible representation $E$ (corresponding to the mDP) and one-dimensional irreducible representations $A_1$ and $A_2$.
Note that each isolated mDP is surrounded by Van Hove singularities in the density of states corresponding to saddle points in the lowest energy minibands. 
The weaker inversion-asymmetric terms, $|\tilde u_i| \ll |u_i|$, in the second line of Eq.~\eqref{eq:H}, open a minigap in both types of mDP discussed above.

Appearance of mDPs at the edge of the first miniband results in a peculiar spectrum of electronic Landau levels, as shown on the r.h.s of Fig.~\ref{fig:fig2}.  
Each data point in these spectra represents one of the Hofstadter minibands \cite{hofstadter_prb_1976} (with an indistinguishably small width) calculated for rational values of magnetic flux, $\frac{p}{q}\Phi_0$ per moir\'e supercell following a method in Ref.~\cite{bistritzer_prb_2011}.
Using these spectra one can trace a clearly separated ``zero-energy'' Landau level related to the isolated $\vect \kappa$-point mDP in the valence band in the bottom row of Fig.~\ref{fig:fig2}, in addition to the true zero-energy Landau level at the conduction-valence band edge.
The three isolated mDPs on the sBZ edge in the valence band (top row of Fig.~\ref{fig:fig2}) also result in a ``zero-energy'' Landau level, though not as clearly separated and split by the magnetic breakdown occurring  at $\Phi\approx0.1\Phi_0$.

%%%%%%%%%%%%%%%%%%%%%%%%%%%%%%%%%%%% conclusions%%%%%%%%%%%%%%%%%%%%%%%%
To summarise, the inversion-symmetric moir\'e perturbation will result in either the first sBZ separated from the rest of the spectrum by one or three mDPs, or, for weak perturbations, will result in overlapping first and higher minibands.
The experimental consequences of this can be expected in the optical spectroscopy of graphene on a hexagonal substrate: the presence of mDPs and Van Hove singularities in the density of states should lead to a modulation of the FIR and IR absorption spectra of monolayer graphene, which otherwise, has the flat absorption coefficient of 2.3\%. 

%  
%%%% Figure: fig3  %%%%%%%%%%%%%%%%%%%%
	\begin{figure}[t]
	\centering
	\includegraphics[width=0.48\textwidth]{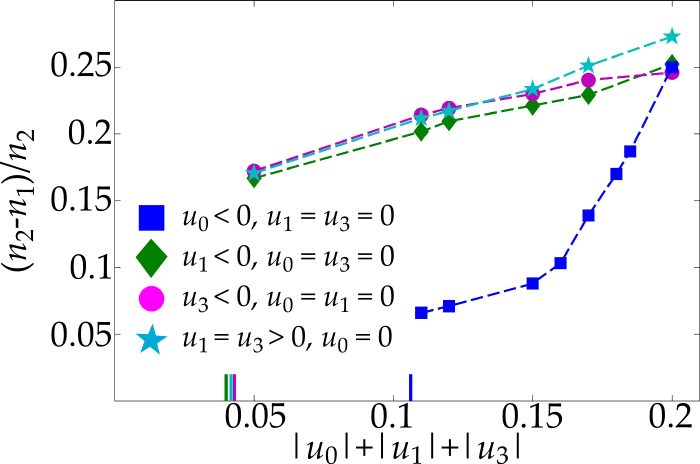}
	\caption{
	The relation between the two densities at which the Hall coefficient in graphene reverses sign upon its doping with holes. The results are shown for several realisations of moir\'e superlattice in the parameter range corresponding to either three isolated mDPs on the sBZ edge (squares) or one isolated mDP at the sBZ corner (other symbols). The thresholds for isolation are indicated on the x-axis.
	}
	\label{fig:fig3}
	\end {figure}

Another experimental consequence of the moir\'e minibands would consist in a non-monotonic variation of the Hall  coefficient upon doping the graphene flake with electrons or holes. For example, for those miniband spectra in Fig.~\ref{fig:fig2}, where there are isolated mDPs in the valence band, the Hall coefficient would pass through a zero value and change sign at two characteristic densities, $n_1$ and $n_2$. At the density $n_1$, which corresponds to the valence band filled with holes up to the Van Hove singularity, the Hall coefficient will change sign from positive to negative. At the higher density, $n_2$, which corresponds to a completely filled first miniband, it would repeat the behaviour at the neutrality point changing sign from negative to positive. Such behaviour is expected to take place for the entire regions of the parametric space painted red, blue or green in Fig.~\ref{fig:fig1}.
The relation between these two carrier densities for various types and strengths of moir\'e perturbations is shown in Fig.~\ref{fig:fig3}.
For the clear part of the parametric space for which we find substantial overlap between many moir\'e minibands such alternations in the sign of the Hall coefficient would be obscured by the competing contributions from the ``electron-like'' and ``hole-like'' branches in the spectrum. 

%%%%%%%%%%%%%%%%%%%%%%%%%%%%%%%%% Acknowledgements %%%%%%%%%%%%%%%%%%%%%%%%%%%%
The authors thank F.~Guinea, A.~MacDonald, E.~J.~Mele and P.~San-Jose for useful discussions during the 2012 KITP programme \emph{The Physics of Graphene}, where this study was started. We acknowledge financial support from DTC NOWNANO, ERC Advanced Grant \emph{Graphene and Beyond}, EU STREP \emph{ConceptGraphene}, Royal Society Wolfson Research Merit Award and EPSRC Science and Innovation Award.\\

%%%%%%%%%%%%%%%%%%%%%%%%%%%%%%%%%%% Appendix 1 %%%%%%%%%%%%%%%%%%%%%%%%%%%%%%%%%%

\section*{Appendix 1: Moir\'e superlattice symmetry}
The point group symmetry of graphene on an incommensurate substrate is given by the intersection of the point group of graphene, $c_{6v}$, with that of its substrate. For a perfectly aligned ($\theta=0$) inversion-symmetric substrate, with either a single (dominant) atom per unit cell or two identical atoms arranged in a honeycomb lattice, the point group symmetries of the substrate and graphene coincide.
The corresponding Hamiltonian, Eq.~\eqref{eq:H}, with moir\'e harmonics orientated as per Fig.~\ref{fig:fig1} (a), must necessarily commute with the operators corresponding to the elements of $c_{6v}$:  $\hat c_6$ , $\hat s_x$ and  $\hat s_y$ which describe $2\pi/6$ rotations and reflections that either exchange or preserve the graphene sublattices. 
The operators for $\hat c_6$ and $\hat s_y$ involve the valley exchanging matrices $\tau_{1,2}$ resulting in that the symmetry of the Hamiltonian restricted to the $K$ valley, as well as the $K$ valley bandstructure, is reduced to $c_{3v}=\{id,\hat c_3,\hat s_x\}$, where $\hat c_3=\hat c_6^2$ has no intervalley structure. 
Each of the $\tilde u_i$ terms are odd under $\hat c_6$, while the $u_2$ and $\tilde u_2$ terms are odd under $\hat s_y$, so that these terms are forbidden for the perfectly aligned inversion-symmetric system described above.
The point group of substrates with the honeycomb lattice and two non-equivalent atoms per unit cell, such as hBN, only possesses the  $\hat c_3$ and  $\hat s_y$ symmetries which allow inversion-asymmetric parameters $\tilde u_{i=0,1,3}$ to take a finite value.

For a finite misalignment angle, the reflection symmetries of graphene and the substrate do not coincide, and the moir\'e harmonics become  misaligned, by an angle $\phi$, from those in Fig.~\ref{fig:fig1} (a). However, the moir\'e harmonics may be brought back into alignment using the transformation $\hat H(\vect r)\rightarrow e^{i\sigma_3\frac{\phi}{2}}\hat H(\hat R_\phi\vect r)e^{-i\sigma_3\frac{\phi}{2}}$, and the $u_2$ and $\tilde u_2$ terms, which are no longer forbidden, may be gauged away. This procedure restores the reflection symmetries to the Hamiltonian, despite their absence in the geometry of the moir\'e pattern for finite misalignment angle.

The symmetries described above can be used to gain a deeper understanding of the mDPs discussed in the main text. The $K$ valley plane wave states from the three equivalent sBZ corners, $\zeta\vect\kappa_{n=0,1,2}=\zeta\hat R_{2\pi n/3}\vect\kappa$,  which form the basis for $\hat H_{\zeta\vect\kappa}$, Eq.~\eqref{eq:kappa}, transform into each other on application of symmetry operators of  $c_{3v}$.  In the same basis, the symmetry operators acting on $ \hat H_{\zeta\vect\kappa}$ take the form of matrices 
\begin{align}
\Gamma^{\zeta \vect\kappa }(\hat c_3)=\begin{pmatrix}0&0&-1\\-1&0&0\\0&1&0       \end{pmatrix}\!,\; 
\Gamma^{\zeta \vect\kappa }(\hat s_x)=s\zeta\begin{pmatrix}1&0&0\\0&0&1\\0&1&0       \end{pmatrix}.\label{eq:symmetry_matrices}
 \end{align}  
For the inversion-symmetric superlattice perturbation, the singlet eigenstate of $\hat H_{\zeta\vect\kappa}$ is given by $\vect v_s=\frac{1}{\sqrt 3}\left(1,-1,-1\right)$. The action of matrices from Eq.~\eqref{eq:symmetry_matrices} on this state show that it transforms according to the one-dimensional irreducible representations of $c_{3v}$: either $A_1$ for $s\zeta=1$ or $A_2$  for $s\zeta=-1$, indicating evenness or oddness under $\hat s_x$  respectively. Similarly, the doublet states of $\hat H_{\zeta\vect\kappa}$, $\vect v_+=\frac{1}{\sqrt 3}\left(\sqrt2,\frac{1}{\sqrt 2},\frac{1}{\sqrt 2}\right)$ and $\vect v_-=\frac{1}{\sqrt 2}\left(0,1,-1\right)$ transform as the two-dimensional irreducible representation, $E$, and their degeneracy is therefore protected by the $c_{3v}$ symmetry.

The  three anisotropic mDPs  can be understood using the compatibility relations in the group appropriate for the sBZ edge, $c_h=\{id, \hat s_x\}$. This group only supports one-dimensional irreducible representations $A_1$ and $A_2$ with the doublet states reducing as $E=A_1 + A_2$.
For a given band, $s=\pm1$, the split bands at $\vect\kappa$ and $-\vect\kappa$ belong to different irreducible representations of $c_h$ and therefore cannot be joined along the sBZ edge. Instead, if both of these bands are closer to zero energy than the doublet states, they must each be joined to one of the doublet bands at the opposite sBZ corner.  
Thus, along the sBZ edge, a crossing of the split bands is required resulting in the mDPs illustrated in the valence band for the top row of Fig.~\ref{fig:fig2}. 

%%%%%%%%%%%%%%%%%%%%%%%%%%%%%%%%%%% Appendix 2 %%%%%%%%%%%%%%%%%%%%%%%%%%%%%%%%%%

\section*{Appendix 2: Microscopic Models}
\subsection{Point charge lattice model}

The point charge model analysed in this Appendix mimics the effect of the quadrupole electric moment of the atoms in the top layer of the substrate. In application to the graphene-hBN system, we neglect the potentials of the quadrupole moments of the boron atom, which have only $\sigma$-orbitals occupied by electrons, and replace nitrogen sites by a point core charge $+2|\text e|$ compensated by the spread out cloud of the $\pi$-electrons, which we replace by a homogeneous background charge density, giving $-2|\text e|$ per hexagonal unit cell of the substrate. This model gives an example of an inversion-symmetric moir\'e superlattice.
The matrix elements of the resulting perturbation, taken between sublattice Bloch states $i$ and $j$ ($i,j=A\text{ or }B$), acting on the low energy Dirac spinors of the graphene $K$ valley, are given by the long wavelength components of 
\begin{widetext}
\begin{align}
  H_{ij}&=\frac{-2 
 \text{e}^2}{4\pi\epsilon_0}\sum_{\vect R_{N}}\int\! d\!z \frac{L^2\Phi^*_{Ki}(\vect r,z)\Phi_{Kj}(\vect r,z)}{\sqrt{(\vect r-\vect R_{N})^2+(z-d)^2}}
 =\!\frac{-2 \text{e}^2}{4\pi\epsilon_0 a}\sum_{\vect g, \vect g',\vect g_{N}}\! I_{|\vect K+\vect g|,|\vect K+\vect g'|,|\vect g_N|}  e^{i(\vect g'-\vect g+\vect g_{N})\cdot\vect r}  e^{i(\!\vect g \cdot \vect \delta_i-\vect g' \cdot \vect \delta_j\!)}.\label{eq:H_charge_model}
\end{align}
\end{widetext}
In Eq.~\eqref{eq:H_charge_model}  $\vect R_N$ are positions of nitrogen sites and $L^2$ is the total area of the graphene sheet; $\Phi_{K,i}(\vect r,z)$  are Bloch wavefunctions of graphene $\pi$-electrons exactly at the $K$ point. Then the Fourier transform has been used to write $\delta \!H_{i j}$ in terms of a sum over substrate reciprocal lattice vectors, $\vect g_N$, and graphene reciprocal lattice vectors, $\vect g$ and $\vect g'$. Nearest neighbour vectors, $\vect \delta_{i=A/B}$, are  $\vect \delta_A=(0,a)$ and $\vect \delta_B=(0,-a)$, so that $\vect K\!\cdot \!\vect \delta_i=0$.
The homogeneous background charge has not been included in Eq.~\eqref{eq:H_charge_model} since its only role is to exclude $\vect g_N=0$ from the sum. The long wavelength terms in the first exponential of the second line of Eq.~\eqref{eq:H_charge_model} determine $\vect b_m=-(\vect g'-\vect g+\vect g_{N})$. The dimensionless integral,
\begin{align*}
&I_{Q,Q', g_N} =\frac{32 a_0^3}{27a^3}\int\! dq_z dq'_z\frac{ \psi^*(Q,q_z)e^{i(q_z-q'_z)\cdot d} \psi(Q',q'_z)}{g_{N}^2+(q_z-q'_z)^2},
\end{align*}
is written in terms of the Fourier transform of the hydrogen-like graphene $\text{P}^{\text{z}}(\vect r,z)$ orbitals with an effective Bohr radius $a_0$,
\begin{align}
\psi(Q,q_z)&=\frac{\pi}{a_0^{3/2}}\frac{1}{2\pi}\int\!d\vect r dz  e^{-i(\vect Q \cdot \vect r+q_z z)} \text{P}^{\text{z}}(\vect r,z)\nonumber\\
&=\frac{-64 i a_0  q_z}{ (1+4 a_0^2 (Q^2+q_z^2))^3}. \nonumber
\end{align}
The integral, $I_{Q,Q',g_N}$, rapidly decays as a function of the magnitude all its arguments so that we limit the sum in Eq.~\eqref{eq:H_charge_model} to only several terms such that $|\vect K+\vect g|=|\vect K+\vect g'|=|\vect K|$, with
$I = I_{K,K,g_0}$ where $g_0=\frac{4\pi}{3a(1+\delta)}$. 

\begin{figure}[htbp]
	\centering
	\includegraphics[width=0.45\textwidth]{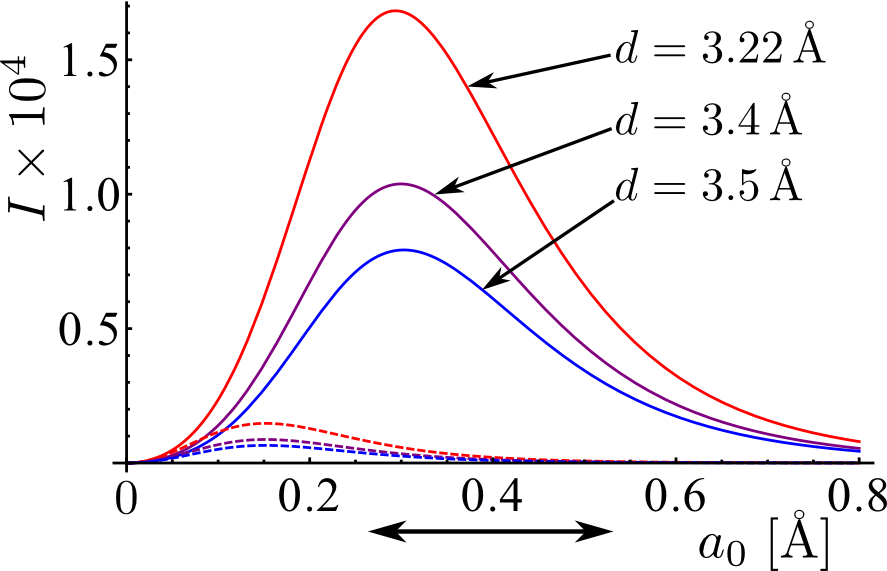}
	\caption{Solid lines show the dimensionless integral $I$, as a function of the effective Bohr radius of the graphene P$^{\text{z}}$ orbitals, for various choices of interlayer separation $d$. 
	To demonstrate convergence of the sum in Eq.~\eqref{eq:H_charge_model}, dashed lines show $I_{2K,K,g_0}$  for the same values of $d$.}
	\label{fig:fig4}
	\end {figure}

For the graphene layer, $a\!=\!1.42\,\text{\AA}$ and for the graphene-hBN heterostructure $\delta\!=\!0.018$. The carbon P$^\text{z}$ orbitals may have a different effective Bohr radius compared to hydrogen.
The range of values quoted  for $\tilde v= \frac{2  \text{e}^2}{4\pi\epsilon_0 a} I$ in Table \ref{table_1} corresponds to the interval $0.27\,\text{\AA}\leq a_0\leq0.53\,\text{\AA}$, indicated by the black double-arrow in Fig.~\ref{fig:fig4}. Interlayer separation   $3.22\,\text{\AA}\leq d\leq 3.5\,\text{\AA}$ is taken from Ref.~\cite{giovannetti_prb_2007}.

Both the dominance of the simplest moir\'e harmonics and the finite values for the off-diagonal terms $u_1$ and $u_2$   stem from the three dimensional treatment of the substrate potential. The potential is strongest near the substrate and therefore a greater proportion of the integral $I_{Q,Q',g_N}$ comes from the region near the substrate, where the graphene P$^\text{z}$ orbitals are broad and therefore have both rapidly decaying Fourier components and significant overlap  with their neighbours. This contrasts with the model employed in Ref. \cite{ortix_prb_2012} which is based on a two-dimensional substrate potential resulting in  $u_1=u_2=0$.

%%%%%%%%%%%%%% Figure: fig4  %%%%%%%%%%%%%%%%%%%%

\subsection{G-hBN hopping model}
In Ref.~\cite{kindermann_prb_2012}, Kindermann \emph{et~al.}~modelled a hBN substrate as a lattice of P$^{\text z}$ orbitals onto which the graphene electrons can hop. This treatment, extended from a model of twisted bilayer graphene \cite{kindermann_prb_2011}, assumed  equal values for the hopping integral to the boron and nitrogen sites, with the difference between the two sublattices arising from their different on-site energies. Here we consider an inversion-symmetric version of the hopping model of Ref.~\cite{kindermann_prb_2012}, assuming that coupling between graphene and the hBN layer is dominated by the hopping to only one of the two sublattices (e.g. boron). Using $\vect k\!\cdot\! \vect p$ theory, this coupling can be written as \cite{kindermann_prb_2012}
\begin{align}
&\delta\!\hat H=\hat H_{\text{int}}\frac{1}{ \epsilon-V-m }\hat H_{\text{int}}^\dagger,\nonumber\\
&\hat H_{int}=\frac{\gamma}{3}\sum_{n=0,1,2} e^{-i \left(\hat R_{ \frac{2\pi n}{3}}\vect \kappa\right)\cdot \vect r} \begin{pmatrix}   e^{i \frac{2\pi n}{3}}\\   e^{-i \frac{2\pi n}{3}} \end{pmatrix}. \label{eq:H_hopping_model}
\end{align}
Neglecting a non-oscillatory term, which corresponds to a trivial constant energy shift, Eq.~\eqref{eq:H_hopping_model} as applied to graphene electrons in valley $K$, leads to the moir\'e Hamiltonian, Eq.~\eqref{eq:H}, with
\begin{align}
\{u_{i=0,...3}\}=\frac{\gamma^2/(vb)}{9(m +V)}\left\{\frac{1}{2}, \frac{- \delta}{\sqrt{\delta^2\!+\!\theta^2}},	  \frac{\theta}{\sqrt{\delta^2\!+\!\theta^2}},- \frac{\sqrt 3}{2}\right\}.\nonumber  
\end{align}
The parameters of the superlattice perturbation given in Table \ref{table_1} of the main text, correspond to $\gamma = 0.3\,\text{eV}$, $V=0.8\,\text{eV}$ and $m=2.3\,\text{eV}$, in accordance with Ref.~\cite{kindermann_prb_2012}.
For the perfectly aligned system, we always find $u_2=0$, which is a consequence of the reflection symmetries present in the perfectly aligned substrate-graphene system (see Appendix 1).


\begin{thebibliography}{99}
\bibitem{footnote_mu_points} The Hamiltonian in the vicinity of other two inequivalent points on the sBZ edge, $\vect\mu'=\hat R_{2\pi/3}\vect\mu$, $\vect\mu''=\hat R_{4\pi/3}\vect\mu$, can be obtained using $H_{\vect\mu+\vect q}=H_{\vect\mu'+\hat R_{2\pi/3}\vect q}=H_{\vect\mu''+\hat R_{4\pi/3}\vect q}$.
\bibitem{footnote_kappa} Note that the spectra derived from $\hat H_{\zeta(\vect\kappa+\vect q)}$ obey the three-fold rotational symmetry.
\end{thebibliography}
\end{document}